\documentclass[jpdc]{apjrnl}
\usepackage{graphics}
\begin{document}

\title{MPICH-G2: A Grid-Enabled Implementation of the Message Passing
       Interface}

\author{Nicholas T. Karonis
\\Department of Computer Science
\\Northern Illinois University
\\DeKalb, IL~~60115
\\Argonne National Laboratory
\\Argonne, IL~~60439
\\Email: karonis@niu.edu
\and
Brian Toonen
\\Mathematics and Computer Science Division
\\Argonne National Laboratory
\\Argonne, IL~~60439
\\Email: toonen@mcs.anl.gov
\and
Ian Foster
\\Argonne National Laboratory
\\Argonne, IL~~60439
\\The University of Chicago
\\Chicago, IL~~60637
\\Email: foster@mcs.anl.gov
}

\date{April 2002}

\maketitle

Proposed running head: MPICH-G2: A Grid-Enabled MPI

\pagebreak

\begin{abstract}

Application development for distributed computing ``Grids'' can
benefit from tools that variously hide or enable application-level
management of critical aspects of the heterogeneous environment. As
part of an investigation of these issues, we have developed
\hbox{MPICH-G2}, a Grid-enabled implementation of the Message Passing
Interface (MPI) that allows a user to run MPI programs across multiple
computers, at the same or different sites, using the same commands that
would be used on a parallel computer.  This library extends the
Argonne MPICH implementation of MPI to use services provided by the
Globus Toolkit for authentication, authorization, resource allocation,
executable staging, and I/O, as well as for process creation,
monitoring, and control.  Various performance-critical operations,
including startup and collective operations, are configured to exploit
network topology information. The library also exploits MPI constructs
for performance management; for example, the MPI communicator
construct is used for application-level discovery of, and
adaptation to, both network topology and network \hbox{quality-of-service}
mechanisms.  We describe the \hbox{MPICH-G2} design and
implementation, present performance results, and review application
experiences, including record-setting distributed simulations.

\end{abstract}


\pagebreak

\section{Introduction}

So-called computational Grids~\cite{GridBook,PhysicsToday} enable the
coupling and coordinated use of geographically distributed resources
for such purposes as large-scale computation, distributed data
analysis, and remote visualization. The development or
adaptation of applications for Grid environments is made challenging,
however, by the often heterogeneous nature of the resources involved and the
facts that that these resources typically live in different
administrative domains, run different software, are subject to
different access control policies, and may be connected by networks
with widely varying performance characteristics.

Such concerns have motivated various explorations of specialized, often
high-level, distributed programming models for Grid environments,
including various forms of object
systems~\cite{grim:legion,GannonChapterCite}, Web
technologies~\cite{FoxChapterCite,webos}, problem solving
environments~\cite{NetSolve,Ninf}, CORBA, workflow systems,
high-throughput computing systems~\cite{Nimrod,Condor}, and
compiler-based systems~\cite{KennedyChapterCite}.

In contrast, we explore here a different approach that might appear
reactionary in its simplicity but that, in fact, delivers a remarkably
sophisticated technology for managing the heterogeneity associated
with Grid environments. Specifically, we advocate the use of a
well-known low-level parallel programming model, the Message Passing
Interface (MPI), as a basis for Grid programming.  While not a
high-level programming model by any means, MPI incorporates
sophisticated support for the management of heterogeneity (e.g., data
types), for the construction of modular programs (the communicator
construct), for management of latency (asynchronous operations), and
for the representation of global operations (collective
operations). These and other features have allowed MPI to achieve
tremendous success as a standard programming model for parallel
computers. We hypothesize that these same features can also be used to
good effect for Grid computing.

Our investigation of MPI as a Grid programming model has focused on
three related questions.  First, can we implement MPI constructs
efficiently in Grid environments to {\em hide} heterogeneity
without introducing overhead?  Second, can we use MPI constructs to
enable users to {\em manage} heterogeneity, when this is required?
Third, do users find MPI useful in practice for application
development?

To allow for the experimental exploration of these questions,
we have developed \hbox{MPICH-G2}, a complete implementation of the
MPI-1 standard~\cite{mpi-forum:journal} that uses services provided by
the Globus Toolkit$^{TM}$~\cite{GlobusHCW98} to extend the popular Argonne
MPICH implementation of MPI~\cite{mpich} for Grid
execution. \hbox{MPICH-G2} passes the MPICH test suite and represents
a complete redesign and reimplementation of the earlier \hbox{MPICH-G}
system~\cite{mpi-nexus-pc} that increases performance significantly 
and incorporates a number of innovations.  
Our experiences with \hbox{MPICH-G2}, as reported in this article,
allow us to respond in the affirmative to each question posed in the
preceding paragraph.

MPICH-G2 hides heterogeneity by using Globus Toolkit services
for such purposes as authentication, authorization, executable
staging, process creation, process monitoring, process control,
communication, redirection of standard input and output, and remote
file access. The result is that a user can run MPI programs across
multiple computers at different sites using the same commands that
would be used on a parallel computer.  Furthermore, performance
studies show that overheads relative to native implementations of
basic communication functions are negligible.

MPICH-G2 enables the use of several different MPI features for
user management of heterogeneity. MPI's asynchronous operations can be
used for latency management in wide-area networks. MPI's communicator
construct can be used to represent the hierarchical structure of
heterogeneous systems and thus allow applications to adapt their
behavior to such structures. (In separate work, we present
topology-aware collective operations as one example of an
``application''~\cite{optcollops}.) We also show how MPI's
communicator construct can be used for user-level management of
network quality of service, as first introduced in an earlier
article~\cite{mpich-gq}.

Many groups have used \hbox{MPICH-G2} for the execution of
both traditional parallel computing applications (e.g., numerical
simulation) and nontraditional distributed computing applications
(e.g., distributed visualization), in both local-area and wide-area
networks. This variety of applications and execution environments
persuades us that MPI can play a valuable role in Grid computing.

MPICH-G2 is not the only implementation of MPI for
heterogeneous systems.  Others include MPICH with the ch\_p4 device
(which provides
limited support for heterogeneity), PACX-MPI~\cite{pacx},
and STAMPI~\cite{stampi}, each of which has
interesting features, as we discuss later. Magpie~\cite{magpie},
IMPI~\cite{impi-web}, and PVM~\cite{pvmbook}
also address relevant issues.
\hbox{MPICH-G2} is
unique, however, in the degree to which it hides and manages heterogeneity, as
well as in its large user community.

In the rest of this article, we describe the problems that we faced in
developing \hbox{MPICH-G2}, the techniques used to overcome these
problems, and experimental results that indicate the performance of
the \hbox{MPICH-G2} implementation and the extent of its improvement
over \hbox{MPICH-G}. We conclude with a discussion of application
experiments and future directions.

\section{Background}

We first provide some brief background on MPI, MPICH, and the Globus
Toolkit.

\subsection{Message Passing Interface}

The Message Passing Interface standard defines a library of
routines that implement the message-passing model. These routines
include {\em point-to-point} communication functions, in which a {\em
send} operation is used to initiate a data transfer between two
concurrently executing program components and a matching {\em receive}
operation is used to extract that data from system data structures
into application memory space; and {\em collective} operations such as
broadcast and reductions that implement operations involving multiple
processes. Numerous other functions address other aspects of message
passing, including, in the MPI-2 extensions to
MPI~\cite{mpi-forum:mpi2-journal}, single-sided communication and
dynamic process creation.

The primary interest of MPI from our perspective, apart from its broad
adoption, is the care taken in its design to ensure that underlying
performance issues are accessible to, not masked from, the programmer.
MPI mechanisms such as asynchronous operations, communicators,
and collective operations all turn out to be useful in Grid environments.

\subsection{MPICH Architecture}

MPICH~\cite{gropp-lusk-doss-skjellum:mpich} is a popular
implementation of the Message Passing Interface standard.  It is a
high-performance, highly portable library originally developed as a
collaborative effort between Argonne National Laboratory and
Mississippi State University.  Argonne continues
research and development efforts aimed at improving MPICH performance
and functionality.

In its present form, MPICH is a complete implementation of the MPI-1
standard with extensions to support the parallel I/O functionality
defined in the MPI-2 standard.  It is a mature, widely distributed
library, with more than 2,000 downloads per month, not including
downloads that occur at mirror sites.  Its free distribution and wide
portability have contributed materially to the adoption of the MPI
standard by the parallel computing community.

MPICH derives its portability from its interfaces and layered
architecture.  At the top is the MPI interface as defined by the MPI
standards.  Directly beneath this interface is the MPICH layer, which
implements the MPI interface.  Much of the code in an MPI
implementation is independent of the networking device or process
management system.  This code, which includes error checking and
various manipulations of the opaque objects, is implemented directly
at the MPICH layer.  All other functionality is passed off to lower
layers be means of the Abstract Device Interface (ADI).

The ADI is a simpler interface than MPI proper
and focuses on moving data between the MPI layer and the network
subsystem.  Those interested in implementing MPI for a particular
platform need only define the routines in the ADI in order to obtain a
full implementation.  Existing implementations of this device
interface for various MPPs, SMPs, and networks provide complete MPI
functionality in a wide variety of environments.  \hbox{MPICH-G2} is
another implementation of the ADI and is otherwise known as the {\em
globus2} device.

\subsection{The Globus Toolkit}

The Globus Toolkit is a collection of software components designed to
support the development of applications for high-performance
distributed computing environments, or
``Grids''~\cite{GlobusHCW98,GridBook}.  
Core components typically define a protocol for interacting with a
remote resource, plus an application program interface (API) used to
invoke that protocol.  (We introduce the protocols and APIs used
within \hbox{MPICH-G2} below.)  Higher-level libraries, services,
tools, and applications use core services to implement more complex
global functionality.
The various Globus Toolkit
components are reviewed in~\cite{Anatomy} and described in detail in
online documentation and in technical papers.

\section{MPICH-G2: A Grid-Enabled MPI}

As noted in the introduction, \hbox{MPICH-G2} is a complete
implementation of the MPI-1 standard that uses Globus Toolkit services
to support efficient and transparent execution in heterogeneous Grid
environments, while also allowing for application management of
heterogeneity.  (It also implements client/server management functions
found in Section 5.4 of the MPI-2
standard~\cite{mpi-forum:mpi2-journal}. However, we do not discuss
these functions here.)

In this section, we first describe the techniques used to hide
heterogeneity during startup and for process management, then the
techniques used to effect communication in heterogeneous systems,
and finally the support provided for application-level management
of heterogeneity.

\subsection{Hiding Heterogeneity during Startup and Management}
\label{sec-startup}

\begin{figure}
\begin{center}
\resizebox{4.5in}{!}{\includegraphics{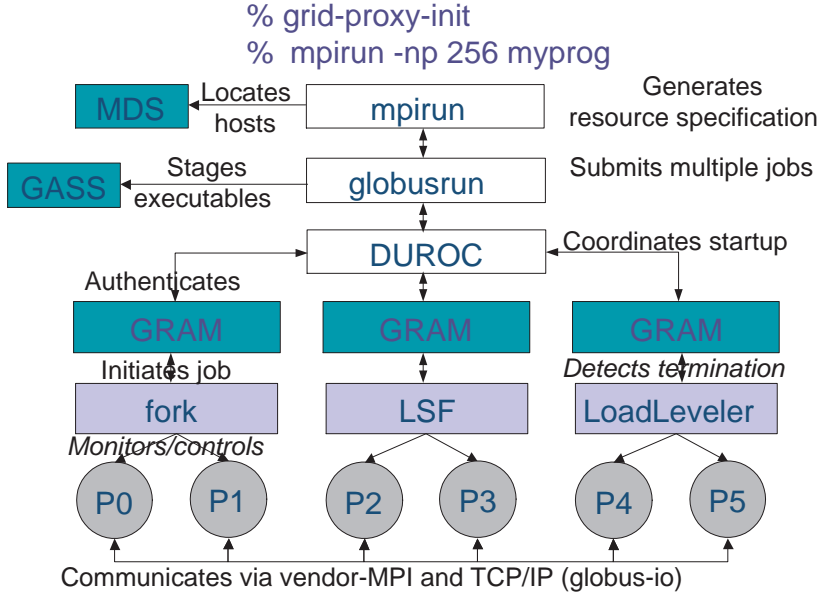}}
\end{center}
\caption{Schematic of the MPICH-G2 startup, showing the various
Globus Toolkit components used to hide and manage heterogeneity. ``Fork,''
``LSF,'' and ``LoadLeveler'' are different local schedulers.}
\label{fig-g}
\end{figure}

As illustrated in Figure~\ref{fig-g} and discussed here,
MPICH-G2 uses a range of Globus Toolkit services to address the
various complex issues that arise in
heterogeneous, multisite Grid environments, such as cross-site
authentication, the need to deal with multiple schedulers with
different characteristics, coordinated process creation, heterogeneous
communication structures, executable staging, and collation of
standard output. In fact, \hbox{MPICH-G2} serves as an exemplary
case study of how Globus Toolkit mechanisms can be used to create
a Grid-enabled programming tool, as we now explain.

Prior to startup of an \hbox{MPICH-G2} application, the user employs
the {\em Grid Security Infrastructure} (GSI)~\cite{GlobusSecurity} to
obtain a (public key) proxy credential that is used to authenticate
the user to each remote sites. This step provides a single sign on
capability.

The user may also use the Monitoring and Discovery Service
(MDS)~\cite{mds97} to select computers on the basis of, for example,
configuration, availability, and network connectivity.

Once authenticated, the user uses the standard {\tt mpirun} command to
request the creation of an MPI computation. The \hbox{MPICH-G2}
implementation of this command uses the {\em Resource Specification 
Language~(RSL)}~\cite{GRAM97} to describe the job.  In brief,
users write {\em RSL scripts}, which identify resources 
(e.g., computers) and specify requirements (e.g., number of CPUs, memory, 
execution time, etc.) and parameters (e.g., location of executables, command 
line arguments, environment variables, etc.) for each.  Based on the
information found in an RSL script, \hbox{MPICH-G2}
calls a {\em co-allocation library}
distributed with the Globus Toolkit, the Dynamically-Updated Request
Online Coallocator (DUROC)~\cite{CoAllocation99}, to schedule and
start the application across the various computers specified by the
user.

The DUROC library itself uses the {\em Grid Resource Allocation and
Management} (GRAM)~\cite{GRAM97} API and protocol to start and
subsequently manage a set of subcomputations, one for each
computer. For each subcomputation, DUROC generates a GRAM request to a
remote GRAM server, which authenticates the user, performs local
authorization, and then interacts with the local scheduler to initiate
the computation. DUROC and associated \hbox{MPICH-G2} libraries tie
the various subcomputations together into a single MPI computation.

GRAM will, if directed, use {\em Global Access to Secondary Storage}
(GASS)~\cite{GASS99} to stage executable(s) from remote locations
(indicated by URLs).  GASS is also used, once an application has
started, to direct standard output and error (stdout and stderr)
streams to the user's terminal, and to provide access to files
regardless of location, thus masking essentially all aspects of
geographical distribution except those associated with performance.

Once the application has started, \hbox{MPICH-G2} selects the most
efficient communication method possible between any two processes,
using vendor-supplied MPI ({\em v}MPI) if available, or {\em Globus
communication} (Globus~IO) with {\em Globus Data Conversion}
(Globus~DC) for TCP, otherwise.

DUROC and GRAM also interact to monitor and manage the execution of the
application.  Each GRAM server monitors the life cycle of its subcomputation
as it passes from pending to running and then to terminating, communicating each
state transition back to DUROC.  Each subcomputation is held at a
DUROC-controlled barrier and is released from that barrier only after 
all subcomputations have started executing.  Also, a request to terminate
the computation (``control C'') may be initiated by the user at which 
time DUROC and the GRAM servers, communicating via GRAM process control
messages, terminate all processes.

\begin{figure}
\begin{center}
\includegraphics{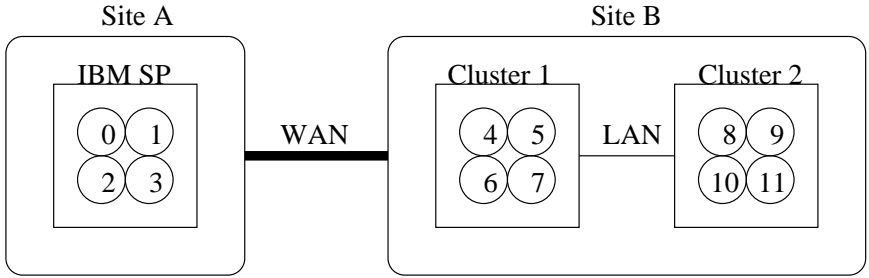}
\end{center}
\caption{An example of an \hbox{MPICH-G2} application running on a 
    computational grid involving 4 processes on an IBM~SP at Site A and 8 
    processes distributed evenly across two Linux clusters at Site B.}
\label{fig-grid}
\end{figure}

\begin{figure}
\begin{center}
\begin{tabular}{|cc|cccccccccccc|}                           \hline
\multicolumn{2}{|c|} {\em Rank}    & 0 & 1 & 2 & 3 & 4 & 5 & 6 & 7 & 8 & 9 & 10 & 11 \\ \hline \hline
\multicolumn{2}{|c|} {\em Depth}   & 4 & 4 & 4 & 4 & 3 & 3 & 3 & 3 & 3 & 3 &  3 &  3 \\ \hline
                     & wide area   & 0 & 0 & 0 & 0 & 0 & 0 & 0 & 0 & 0 & 0 & 0  &  0 \\ 
{\em Colors}         & local area  & 0 & 0 & 0 & 0 & 1 & 1 & 1 & 1 & 1 & 1 &  1 &  1 \\
                     & system area & 0 & 0 & 0 & 0 & 1 & 1 & 1 & 1 & 2 & 2 &  2 &  2 \\
                     & {\em v}MPI  & 0 & 0 & 0 & 0 &   &   &   &   &   &   &    &    \\ \hline
\end{tabular}
\end{center}
\caption{An example of {\em depths} and {\em colors} used by \hbox{MPICH-G2}
    to represent network topology in a computational grid.}
\label{fig-depths-colors}
\end{figure}

After the processes have started, \hbox{MPICH-G2} uses information
specified in the RSL script to create {\em multilevel clustering} of the 
processes based on the underlying network topology.
Figure~\ref{fig-grid}
depicts an MPI application involving 12 processes distributed across 
three machines located at two sites.  We depict 4 processes 
(\verb+MPI_COMM_WORLD+ ranks 0-3) on the IBM~SP at Site~A and 
4 processes on each of two Linux clusters (\verb+MPI_COMM_WORLD+ ranks 4-7
and 8-11, respectively) at Site~B.
Each process in \verb+MPI_COMM_WORLD+ is assigned a {\em topology depth}.
Processes
that communicate using only TCP are assigned topology depths of 3 
(to distinguish between wide area, local area, and intramachine TCP 
messaging), and processes
that can also communicate using a {\em v}MPI have a topology
depth of 4.
Using these topology depths \hbox{MPICH-G2} groups processes at a particular
level through the assignment of {\em colors}.  Two processes are assigned
the same color at a particular level if they can communicate with each
other at the network level.

Figure~\ref{fig-depths-colors} depicts the {\em topology depths}
and {\em colors} for the processes depicted in Figure~\ref{fig-grid}.  Those
processes capable of communicating over {\em v}MPI, (i.e., those executing
on the IBM~SP), have a depth of 4, while the other processes, (i.e., those 
executing on a Linux cluster), have a depth of 3.  Since all processes are on 
the same wide-area network, they all have the same {\em color} (0) at the
wide-area level.  Similarly, at the local-area level, all the processes 
at Site A are assigned one color (0), while all the processes at Site B
are assigned another (1).  This structure continues through the system-area level,
where processes are assigned the same color if and only if they are 
on the same machine. Finally, processes that can communicate
over a {\em v}MPI are assigned the same color at the {\em v}MPI
level if and only if they can communicate directly with each other over
the {\em v}MPI.

Topology depths and colors are used in the multilevel topology-aware
collective operations and topology-discovery mechanism described in
Sections~\ref{g2-impl} and~\ref{g2-mgmt}, respectively.

\subsection{Heterogeneous Communications}
\label{g2-impl}

MPICH-G2 achieves major performance improvements relative to
the earlier \hbox{MPICH-G}~\cite{mpi-nexus-pc} by replacing
Nexus~\cite{JPDCNexus}, the multimethod, single-sided communication library used for
all communication in \hbox{MPICH-G}, with specialized MPICH-specific
communication code.  While Nexus has attractive features (e.g.,
multiprotocol support with highly tuned TCP support and automatic data
conversion), other attributes have proved less
attractive from a performance perspective. \hbox{MPICH-G2} now handles
all communication directly by reimplementing the good things about
Nexus and improving the others.  The result, as we show in
Section~\ref{sec-exp}, is that we achieve performance virtually
identical to vendor MPI and MPICH configured with the default TCP
(ch\_p4) device.  We provide here
a detailed description of the improvements and additions to
\hbox{MPICH-G} used to achieve this impressive performance.

\paragraph{Increased bandwidth.}
In \hbox{MPICH-G}, each communication involved the copying of data
to and from Nexus buffers in sending and receiving processes.
\hbox{MPICH-G2} eliminates these two extra copies in the case of
intramachine messages where a vendor MPI exists. In this situation, 
sends and receives now flow directly from and to application buffers, respectively.  In
addition, for TCP
messaging involving basic MPI datatypes (e.g., \verb+MPI_INT+, 
\verb+MPI_FLOAT+) the sending process also transmits directly 
from the application buffer.

\paragraph{Reduced latency for intramachine vendor MPI messaging.}
Multiprotocol support is achieved in Nexus by polling each protocol
(TCP, vendor MPI, etc.) for incoming messages in a roundrobin
fashion~\cite{MultiMethodJPDC}. However, this strategy is inefficient
in many situations: it is relatively expensive to poll a TCP socket
and in practice it is often the case that many processes in a
\hbox{MPICH-G2} computation use only vendor MPI (for communicating with
other processes on the same machine).

While this inefficiency can be reduced by adaptive
polling~\cite{MultiMethodJPDC} or by introducing distinct
proxy processes~\cite{pacx,stampi}, \hbox{MPICH-G2} takes a more direct
approach, exploiting the knowledge about message source that is
provided by TCP receive commands to eliminate TCP polling altogether
in many situations.  \hbox{MPICH-G2} polls TCP {\em only} when the
application is expecting data from a source that dictates, or might
dictate (e.g., \verb+MPI_Recv+ specifies source=\verb+MPI_ANY_SOURCE+),
TCP messaging.

This avoidance of unnecessary polling when coupled with the need to
guarantee progress on both the vendor MPI and TCP protocols leads to
implementation decisions that can affect an application's
point-to-point communication performance.  Specifically, for processes
executing on machines where a vendor MPI is available, the context in
which the application calls \verb+MPI_Recv+ affects the manner in
which \hbox{MPICH-G2} implements that function, as follows:

\begin{itemize}
\item {\bf Specified.}
The source rank specified in the call to \verb+MPI_Recv+ explicitly
identifies a process on the same machine (in the same vendor MPI job).
Furthermore, no asynchronous requests are outstanding (e.g., incomplete
\verb+MPI_Irecv+ and/or \verb+MPI_Isend+).  If these two conditions
are met, \hbox{MPICH-G2} implements \verb+MPI_Recv+ by directly
calling the \verb+MPI_Recv+ of the underlying vendor MPI.  This is the
most favorable circumstances under which an \verb+MPI_Recv+ can be
performed.

\item {\bf Specified-pending.}
This category is similar to the {\em specified} category in that the
\verb+MPI_Recv+ specifies an explicit source rank on the same machine.
This time, however, one or more unsatisfied receive requests are
present, and each such request specifies a source on the same machine.
This situation forces \hbox{MPICH-G2} to continuously poll
(\verb+MPI_Iprobe+) the vendor MPI for incoming messages.  This
scenario results in less efficient MPICH-G2 performance since the
induced polling loop increases latency.

\item {\bf Multimethod.}
Here the source rank for the \verb+MPI_Recv+ is \verb+MPI_ANY_SOURCE+
or \verb+MPI_Recv+ is called in the presence of unsatisfied
asynchronous requests that require, or might require, TCP messaging.
In this situation, \hbox{MPICH-G2} must poll both TCP and the
vendor MPI continuously.  This is the least efficient \hbox{MPICH-G2} scenario, since the relatively large cost of TCP polling results in even greater
latency.
\end{itemize}
In Section~\ref{sec-exp}, we present a quantitative analysis of the performance differences that
result from these different structures.

\paragraph{More efficient use of sockets.}
The Nexus single-sided communication paradigm results in \hbox{MPICH-G2}
opening {\em two pairs of 
sockets} between communicating processes and using each pair as a 
simplex channel (i.e., data always flowing in one direction over each socket 
pair). \hbox{MPICH-G2} opens a {\em single pair of sockets} between two 
processes and sends data in both directions.  This approach reduces the use 
of system resources; moreover, by using sockets in the bidirectional manner in which 
they were intended, it also improves TCP efficiency.

\paragraph{Multilevel topology-aware collective operations.}
Early implementations of MPI's collective operations sought to
construct communication structures that were optimal under the
assumption that all processes were equidistant from one
another~\cite{postal,logp}. Since this assumption is unlikely to be valid in
Grid environments, however, it is desirable that a Grid-enabled MPI
incorporate collective operation implementations that take into
account the actual topology. \hbox{MPICH-G2} does this, and we have
demonstrated substantial performance improvements for our {\em
multilevel topology-aware} approach~\cite{optcollops} relative both to
topology-{\em un}aware binomial trees and earlier topology-aware
approaches that distinguish only between ``intracluster'' and
``intercluster'' communications~\cite{starT,magpie}.

As we explain in the next subsection, \hbox{MPICH-G2}'s topology-aware
collective operations are constructed in terms of topology discovery
mechanisms that can also be used by topology-aware applications.

\subsection{Application-Level Management of Heterogeneity}
\label{g2-mgmt}

We have experimented within MPICH-G2 with a variety of mechanisms for
application-level management of heterogeneity in the underlying
platform. We mention two here.

\paragraph{Topology discovery.}
Once an MPI program starts, all processes can be viewed as equivalent,
distinguished only by their rank. This level of abstraction is
desirable from a programming viewpoint but makes it difficult to
write programs that exploit aspects of the underlying physical
topology, for example, to minimize expensive intercluster
communications.

MPICH-G2 addresses this issue {\em within the standard MPI framework}
by using the MPI communicator construct to deliver topology
information to an application. 
It associates {\em attributes} with each MPI
communicator to communicate this topology information, which is expressed
within each process in terms of {\em topology depths} and
{\em colors}, as described in Section~\ref{sec-startup}.

\begin{figure}
\begin{small}
\begin{verbatim}
#include <mpi.h>

int main(int argc, char *argv[])
{
  int me, flag;
  int *depths;
  int **colors;
  MPI_Comm LANcomm, VcommA, VcommB;

  MPI_Init(&argc, &argv);
  MPI_Comm_rank(MPI_COMM_WORLD, &me);
  MPI_Attr_get(MPI_COMM_WORLD, MPICHX_TOPOLOGY_DEPTHS, &depths, &flag);
  MPI_Attr_get(MPI_COMM_WORLD, MPICHX_TOPOLOGY_COLORS, &colors, &flag);

  MPI_Comm_split(MPI_COMM_WORLD, colors[me][1], 0, &LANcomm);
  MPI_Comm_split(MPI_COMM_WORLD, (depths[me] == 4 ? colors[me][3] : -1), 
                 0, &VcommA);
  MPI_Comm_split(MPI_COMM_WORLD, 
                 (depths[me] == 4 ? colors[me][3] : MPI_UNDEFINED), 
                 0, &VcommB);

  MPI_Finalize();
}
\end{verbatim}
\end{small}
\caption{An example \hbox{MPICH-G2} application that uses {\em topology depths}
    and {\em colors} to create communicators that group processes into
    various topology-aware clusters.}
\label{fig-hier}
\end{figure}

MPICH-G2 applications can then query communicators to retrieve
attribute values and structure themselves appropriately. For example,
it is straightforward to create new communicators that reflect
the underlying network topology. Figure~\ref{fig-hier}
depicts an \hbox{MPICH-G2} application that first queries
the \hbox{MPICH-G2-defined} communicator attributes 
\verb+MPICHX_TOPOLOGY_DEPTHS+ and \verb+MPICHX_TOPOLOGY_COLORS+
to discover topology depths and colors, respectively, and
then uses those values to create three communicators: \verb+LANcomm+,
which groups processes based on site boundaries, \verb+VcommA+, which
groups processes based on their ability to communicate with each
other over {\em v}MPI, while placing all processes that cannot communicate
over {\em v}MPI into a separate communicator, and \verb+VcommB+,
which groups the processes in much the same way as \verb+VcommA+, but 
this time does not place processes that cannot communicate over {\em v}MPI 
in a communicator (i.e., \verb+VcommB+ is set to 
\verb+MPI_COMM_NULL+ for those processes).  

\paragraph{Quality-of-service management.}
We have experimented with similar techniques for purposes of quality of
service management~\cite{mpich-gq}. When running over a shared
network, an MPI application may wish to negotiate with an external
resource management system to obtain dedicated access to (part of) the
network. We show that communicator attributes can be used to set and
initiate \hbox{quality-of-service} parameters between selected processes.

\section{Performance Experiments}
\label{sec-exp}

We present the results of detailed performance experiments that
characterize the performance of \hbox{MPICH-G2} and demonstrate the
major improvements achieved relative to its predecessor,
\hbox{MPICH-G}.  We begin by looking at the performance of {\em
intramachine} communication over a vendor MPI.  Then, we examine
performance when TCP is the only choice for communicating between a
pair of processes.  In all cases, mpptest~\cite{pvmmpi99-mpptest},
the performance tool included in the MPICH distribution, is used to
obtain all results.

\subsection{Vendor MPI}

Evaluating the performance of MPICH-G2 when using a vendor MPI as an
underlying communication mechanism is not as simple as running a
single set of ping-pong tests.  As discussed earlier, the performance
achieved by \hbox{MPICH-G2} can be affected by outstanding requests
and by the use of \verb+MPI_ANY_SOURCE+.  Therefore, we have divided the
experiments into the three categories described in Section~\ref{g2-impl}.

Our vendor MPI experiments were run on an SGI Origin2000 at Argonne
National Laboratory.
Both \hbox{MPICH-G2} and \hbox{MPICH-G} were built using a
nonthreaded, no-debug flavor of Globus 1.1.4 and performed
intramachine communication via SGI's implementation of MPI.

\begin{figure}
\begin{center}
\includegraphics{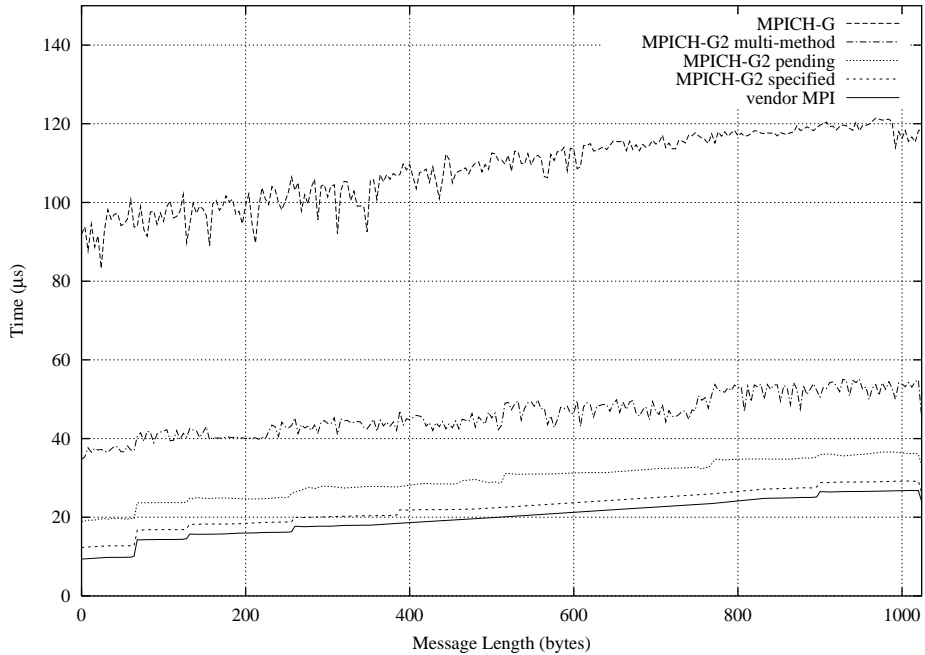}
\end{center}
\caption{vMPI experiments -- small message latency.}
\label{fig-vmpi-lat}
\end{figure}
One \hbox{MPICH-G2} design goal was to minimize latency overhead for
intramachine communication relative to an underlying vendor MPI.  As
can been seen in Figure~\ref{fig-vmpi-lat}, \hbox{MPICH-G2} does an
outstanding job in this regard: only a few extra microseconds of
latency are introduced by \hbox{MPICH-G2} when the source of the
message is specified and no other requests are outstanding.  In
contrast, \hbox{MPICH-G} added approximately 80 microseconds of
latency to each message, because the multiple steps required to
implement the Nexus single-sided communication model.

The introduction of pending receive requests has a modest impact on
\hbox{MPICH-G2} message latencies.  Messages falling into the {\em
specified-pending} category incur slightly more overhead, as the
\hbox{MPICH-G2} progress engine must continuously poll (probe) the
vendor MPI rather than blocking in a receive.  Overall,
\hbox{MPICH-G2} latencies increase by several microseconds relative to
the first case but are still far less than those of \hbox{MPICH-G}.

The use of \verb+MPI_ANY_SOURCE+ has the largest impact on
\hbox{MPICH-G2} performance.  The additional cost is associated with
having to poll TCP as well as the vendor MPI.  Polling TCP
increases the latency of
messages by nearly 20 microseconds over those in the {\em
specified-pending} category.  While the increase is significant, however, these
latencies are still considerably less than for \hbox{MPICH-G}.

\begin{figure}
\begin{center}
\includegraphics{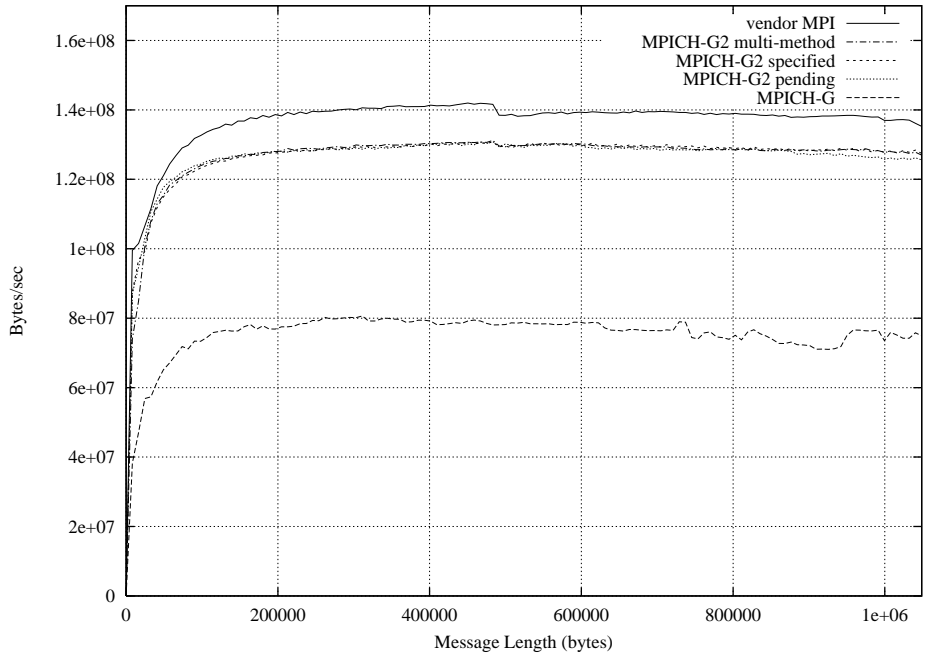}
\end{center}
\caption{vMPI experiments -- realized bandwidth.}
\label{fig-vmpi-bw}
\end{figure}
While \hbox{MPICH-G2} message latencies are affected by the use of
\verb+MPI_ANY_SOURCE+ and pending receive requests, the realized
bandwidths are largely unaffected.  Figure~\ref{fig-vmpi-bw} shows
the bandwidths obtained for messages up to one megabyte.
We see that the bandwidths for
\hbox{MPICH-G2} are nearly identical for all but small
messages.  While the large message bandwidths for \hbox{MPICH-G2} are
approximately 7\% less than those for the the vendor MPI (for reasons
we do not yet understand), they
represent an improvement of more than 60\% over \hbox{MPICH-G}.

\subsection{TCP/IP}

Performance optimization work on \hbox{MPICH-G2} performed to date has
focused on intramachine messaging when a vendor MPI is used as the
underlying communication mechanism.  The \hbox{MPICH-G2} TCP/IP
communication code has not been optimized.  However, its performance
is quite reasonable when compared with \hbox{MPICH-G} and to MPICH
configured with the default TCP (ch\_p4) device.

All TCP/IP performance measurements were taken using a pair of
SUN workstations in Argonne's Mathematics and Computer Science
Division.  These two machines were connected to a local-area network
via gigabit Ethernet.  Both \hbox{MPICH-G} and \hbox{MPICH-G2} were
built using a nonthreaded, no-debug flavor of Globus 1.1.4.

\begin{figure}
\begin{center}
\includegraphics{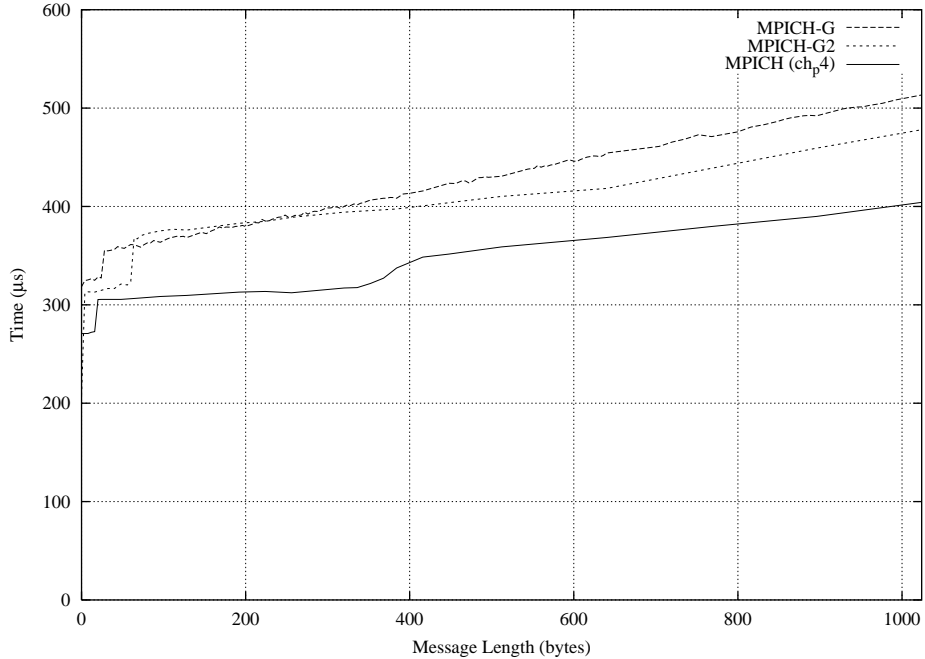}
\end{center}
\caption{TCP/IP experiments -- small message latency.}
\label{fig-tcp-lat}
\end{figure}
Figure~\ref{fig-tcp-lat} shows the small message
latencies exhibited by all three systems.  We see that
for most message sizes, \hbox{MPICH-G2} is 20\% to 30\% slower than
MPICH/ch\_p4, although the difference is much smaller for very small
messages.  We also see that \hbox{MPICH-G2} latencies, in most cases,
are somewhat less than those of \hbox{MPICH-G}.

The most notable data point is barely visible on the graph but
emphasizes a clear optimization that is missing in \hbox{MPICH-G2}.
The latency for zero-byte messages is 140 microseconds, while the
latency for an eight-byte message is 224 microseconds.  The reason for
this large difference is that \hbox{MPICH-G2} currently uses separate
system calls to send
the message header and the message data.
This data point suggests that
by combining these two writes into a single vector write, we could
reduce the latency of small messages significantly.  While this
difference might seem unimportant for machines separated by a wide-area
network, it can be significant when \hbox{MPICH-G2} is used to
combine multiple machines with the same machine room or even at the
same site.

\begin{figure}
\begin{center}
\includegraphics{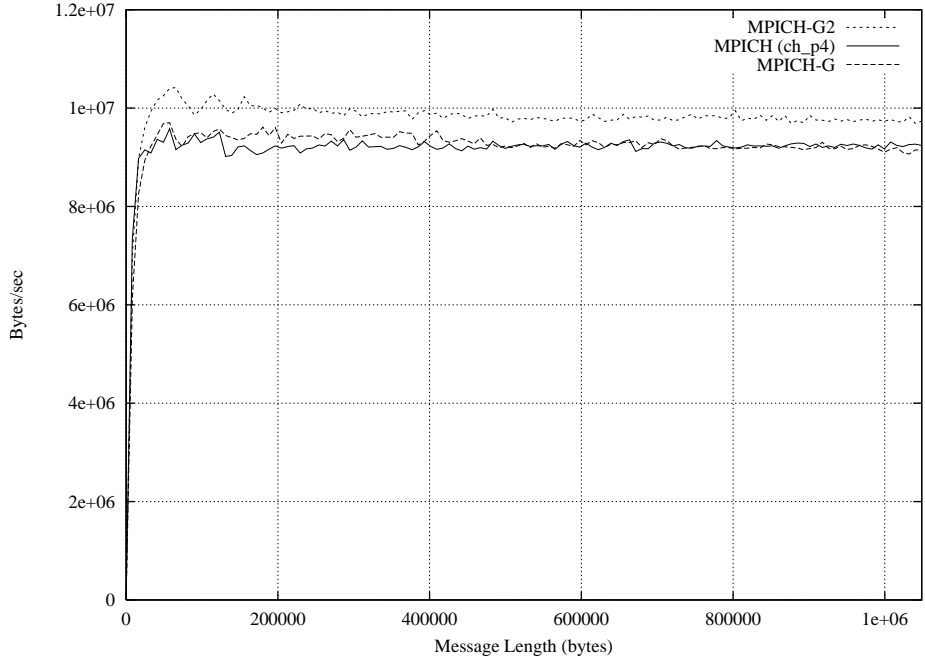}
\end{center}
\caption{TCP/IP experiments -- realized bandwidth.}
\label{fig-tcp-bw}
\end{figure}

Figure~\ref{fig-tcp-bw} shows the bandwidths
obtained by all three systems for message sizes up to one megabyte.
For large messages, we see that \hbox{MPICH-G2} performs approximately
5\% better than the other two systems.  This improvement is a result
of the message data being sent directly from the user buffer rather
than being copied into a separate buffer before \verb+write+ is
called.  For preposted receives with contiguous data, further
improvement is possible.  Data for these receives can be read directly
into the user buffer, avoiding a buffer copy that, at present, always
takes place at the receiver.

\section{Application Experiences}
\label{sec-apps}

MPICH-G2 has been used by many groups worldwide for a wide variety
of purposes. Here we mention a few relevant experiences that
highlight interesting features of the system.

One interesting use of MPICH-G2 is to run conventional MPI programs
across multiple parallel computers within the same machine room.  In
this case, \hbox{MPICH-G2} is used primarily to manage startup and to
achieve efficient communication via use of different low-level
communication methods.  Other groups are using \hbox{MPICH-G2} to
distribute applications across computers located at different sites,
for example, Taylor performing MM5 climate modeling on the NSF
TeraGrid~\cite{teragrid,ncsa-news}, Mahinthakumar forming multivariate
geographic clusters to produce maps of regions of ecological
similarity~\cite{kumarsc99}, Larsson for studies of distributed
execution of a large computational electromagnetics code~\cite{Olle},
and Chen and Taylor in studies of automatic partitioning techniques,
as applied to finite element codes~\cite{Jian}.

MPICH-G2 has also been successfully used in demonstrations that
promote MPI as an application-level interface to Grids for
nontraditional distributed computing applications, for example, Roy et al.\
for studies in using MPI idioms for setting QoS
parameters~\cite{mpich-gq} and Papka and Binns for creating
distributed visualization pipelines using \hbox{MPICH-G2's} client/server MPI-2
extensions~\cite{teragrid,ncsa-news}.

\hbox{MPICH-G2} was awarded a 2001 Gordon Bell Award
for its role in an astrophysics application used for solving problems
in numerical relativity to study gravitational waves from colliding
black holes~\cite{CactusGBsc01}.  The winning team used
\hbox{MPICH-G2} to run across four supercomputers in California and
Illinois, achieving scaling of 88\% (1,140 CPUs) and 63\% (1,500 CPUs)
computing a problem size five times larger than any other previous
run.

\section{Future Work}

The successful development of MPICH-G2 and its widespread adoption
both make it a useful platform for future research and create
significant interest in its continued development.

One immediate area of concern is full support for MPI-2 features.  In
particular, support for dynamic process management will allow
\hbox{MPICH-G2} to be used for a wider class of Grid computations in
which either application requirements or resource availability changes
dynamically over time. The necessary support exists in the Globus
Toolkit, and so this work depends primarily on the availability of the
next-generation ADI-3. Less obvious, but very interesting, is how to
integrate support for fault tolerance into \hbox{MPICH-G2} in a
meaningful way.

A second area of concern relates to exploring and refining
\hbox{MPICH-G2} support for application-level management of
heterogeneity. Initial experiments with topology discovery and 
quality-of-service management have been encouraging, but it seems inevitable
that application experiences will reveal deficiencies in current
techniques or suggest additional \hbox{MPICH-G2} support that
could further improve application flexibility.

Our work on collective operations can be improved in various ways.  In
particular, van de Geijn et al.~\cite{pipelining} have shown that are
advantages in implementing collective operations by segmenting and
pipelining messages when communicating over relatively slower
channels (e.g., TCP over local- and wide-area networks).  These
pipelining techniques can be used throughout many of the levels in
\hbox{MPICH-G2's} multilevel topology-aware collective operations.

\section{Related Work}

A variety of approaches have been proposed to programming Grid
applications, including object
systems~\cite{grim:legion,GannonChapterCite}, Web
technologies~\cite{FoxChapterCite,webos}, problem solving
environments~\cite{NetSolve,Ninf}, CORBA, workflow systems,
high-throughput computing systems~\cite{Nimrod,Condor}, and
compiler-based systems~\cite{KennedyChapterCite}. We assume that while
different technologies will prove attractive for different purposes, a
programming model such as MPI that allows direct control over low-level
communications will always be attractive for certain applications.

Other systems that support message passing in heterogeneous environments
include the pioneering Parallel Virtual Machine
(PVM)~\cite{pvmbook,PVMATM2} and the PACX-MPI~\cite{pacx},
MetaMPI~\cite{metampi}, and STAMPI~\cite{stampi} implementations of MPI,
each of which addresses issues relating to efficient communication in
heterogeneous wide-area systems.  STAMPI supports MPI-2 dynamic process
management features.  PACX-MPI, like \hbox{MPICH-G2}, supports the
automatic startup of distributed computations, but uses ssh rather than 
the GRAM protocol with its integrated
GSI authentication, for that 
purpose; nor does it address issues of executable staging. 
PACX-MPI (and STAMPI) also differ in how it
addresses wide-area communication. While in \hbox{MPICH-G2}, any
processor may speak both local and wide-area communication protocols,
PACX-MPI and STAMPI forward all off-cluster communication operations to
an intermediate gateway node.

Other implementations of MPI include MPICH with the ch\_p4 device
and LAM/MPI~\cite{lam,lam-www}.  By contrast these implementations
were designed for local area networks and not computational grids.

The Interoperable MPI (IMPI) standards effort~\cite{impi-web} defines
standard message formats and protocols with a view to enabling
interoperability among different MPI implementations. IMPI does {\em
not} address issues of computation management and control;
in principle, the techniques developed within \hbox{MPICH-G2} could be
used for that purpose.

Other related projects include MagPIe~\cite{magpie} and
MPI-StarT~\cite{starT}, which show how careful consideration of
communication topologies can result in significant improvements after
modifying the MPICH broadcast algorithm, which uses topology-{\em
un}aware binomial trees.  However, both limit their view of the
network to only two layers; processors are either near or far.
Further performance improvements can be realized by adopting the
multilevel network view.  We referred in the preceding section to the
work of van de Geijn et al.~\cite{pipelining}.  In~\cite{magpie-PC}
Kielman et al. have extended MagPIe by incorporating van de Geijn's
pipelining idea through a technique they call Parameterized LogP
(PLogP), which is an extension of the LogP model presented
by Culler et al~\cite{logp}.  In this extension, MagPIe still recognizes only a
two-layer communication network, but through parameterized studies of
the network they determine ``optimal'' packet sizes.

Various projects have investigated programming model extensions to
enable application management of QoS, for example, Quo~\cite{Quo}.
The only other relevant effort in the context of MPI is work on
real-time extensions to MPI.  MPI/RT~\cite{mpirt-web} provides a QoS
interface but is not an established standard and introduces a new
programming interface.  Furthermore, the focus is on real-time needs
such as predictability of performance and system resource usage more
appropriate for embedded systems than for wide-area networks.

\section{Summary}

We have described \hbox{MPICH-G2}, an implementation of the Message
Passing Interface that uses Globus Toolkit mechanisms to support the
execution of MPI programs in heterogeneous wide-area
environments. \hbox{MPICH-G2} masks details of underlying networks,
software systems, policies, and computer architectures so that diverse
distributed resources can appear as a single {\tt
MPI\_COMM\_WORLD}. Arbitrary MPI applications can be started on
heterogeneous collections of machines simply by typing mpirun:
authentication, authorization, executable staging, resource
allocation, job creation, startup, and routing of stdout and stderr
are all handled automatically via Globus Toolkit mechanisms.
\hbox{MPICH-G2} also enables the use of MPI features for user-level
management of heterogeneity, for example, via the use of MPI's
communicator construct to access system topology information.  A wide
range of successful application experiences have demonstrated
\hbox{MPICH-G2}'s utility in practical settings, both for traditional
simulation applications and for less traditional applications such as
distributed visualization pipelines.

While \hbox{MPICH-G2} is already a sophisticated tool that is seeing
widespread use, there are also several areas in which it can be
extended and improved. Support for MPI-2 features, in particular
dynamic process management, will be invaluable for Grid applications
that adapt their resource usage to changing conditions and application
requirements.  This support will be provided as soon as it is
incorporated into MPICH. More challenging is the design of techniques
for effective fault management, a major topic for future research.
Here we may be able to draw upon techniques developed within systems
such as PVM~\cite{pvmbook}.

\begin{acknowledge}

We thank Olle~Larsson and Warren~Smith for early discussions and for
prototyping the techniques that enable us to use vendor-supplied MPI.
\hbox{MPICH-G2} is, to a large extent, the result of our
\hbox{MPICH-G} experiences.  We therefore thank Jonathan~Geisler, who
originally designed and implemented \hbox{MPICH-G} while at Argonne,
and George~Thiruvathukal, who further developed \hbox{MPICH-G} also
while at Argonne.  We thank William~Gropp, Ewing~Lusk, David~Ashton,
Anthony~Chan, Rob~Ross, Debbie~Swider, and Rajeev~Thakur of the MPICH
group at Argonne for their guidance, assistance, insight, and many
discussions.  We thank Sebastien~Lacour for his efforts in conducting
the performance evaluation and his many other contributions. His
insight and ingenuity were invaluable to the implementation of the
topology-aware components of \hbox{MPICH-G2}.  Finally, we thank all
the members of the Globus development team for their support,
patience, and many ideas.

This work was supported in part by the Mathematical, Information, and
Computational Sciences Division subprogram of the Office of Advanced
Scientific Computing Research, U.S. Department of Energy, under Contract
W-31-109-Eng-38; by the U.S. Department of Energy under Cooperative
Agreement No. DE-FC02-99ER25398; by the Defense Advanced Research
Projects Agency under contract N66001-96-C-8523; by the National Science
Foundation; and by the NASA Information Power Grid program.

\end{acknowledge}

\bibliographystyle{plain}
\bibliography{foster_bibliography,mpi-allbib,mpi-book,globus,prop}

\begin{thebibliography}{10}

\bibitem{Nimrod}
D.~Abramson, R.~Sosic, J.~Giddy, and B.~Hall.
\newblock Nimrod: A tool for performing parameterised simulations using
  distributed workstations.
\newblock In {\em Proc. 4th IEEE Symp. on High Performance Distributed
  Computing}. IEEE Computer Society Press, 1995.

\bibitem{CactusGBsc01}
G.~Allen, T.~Dramlitsch, I.~Foster, M.~Ripeanu N.~T.~Karonis, E.~Seidel, and
  B.~Toonen.
\newblock Supporting efficient execution in heterogeneous distributed computing
  environments with catus and globus.
\newblock In {\em Proceedings of Supercomputing 2001}. IEEE Computer Society
  Press, 2001, winner Gordon Bell Award, Special Category.

\bibitem{pipelining}
M.~Barnett, R.~Littlefield, D.~Payne, and R.~van~de Geijn.
\newblock On the efficiency of global combine algorithms for 2-d meshes with
  wormhole routing.
\newblock {\em Journal of Parallel and Distributed Computing}, 22:324--328,
  1994.

\bibitem{postal}
A.~Bary-Noy and S.~Kipnis.
\newblock Designing broadcasting algorithms in the postal model for
  message-passing systems.
\newblock In {\em Proceedings of the 4th Annual ACM Symposium on Parallel
  Algorithms and Architectures}, pages 559--566, June 1992.

\bibitem{GASS99}
Joseph Bester, Ian Foster, Carl Kesselman, Jean Tedesco, and Steven Tuecke.
\newblock {GASS}: {A} data movement and access service for wide area computing
  systems.
\newblock In {\em Proc.\ IOPADS'99}. ACM Press, 1999.

\bibitem{lam}
Greg Burns, Raja Daoud, and James Vaigl.
\newblock {LAM}: An open cluster environment for {MPI}.
\newblock In John~W. Ross, editor, {\em Proceedings of Supercomputing Symposium
  '94}, pages 379--386. University of Toronto, 1994.

\bibitem{NetSolve}
Henri Casanova and Jack Dongarra.
\newblock Netsolve: A network server for solving computational science
  problems.
\newblock Technical Report CS-95-313, University of Tennessee, November 1995.

\bibitem{Jian}
Jian Chen and Valerie Taylor.
\newblock Mesh partitioning for distributed systems.
\newblock In {\em Proc. 7th IEEE Symp. on High Performance Distributed
  Computing}. IEEE Computer Society Press, 1998.

\bibitem{logp}
D.E. Culler, R.~Karp, D.A. Patterson, A.~Sahay.~K.E. Schauser, E.~Santos,
  R.~Subramonian, and T.~von Eicken.
\newblock Logp: Towards a realistic model of parallel compuation.
\newblock In {\em Proceedings of the 4th SIGPLAN Symposium on Principles and
  Practices of Parallel Programming}, pages 1--12, May 1993.

\bibitem{GRAM97}
K.~Czajkowski, I.~Foster, N.~Karonis, C.~Kesselman, S.~Martin, W.~Smith, and
  S.~Tuecke.
\newblock A resource management architecture for metacomputing systems.
\newblock In {\em The 4th Workshop on Job Scheduling Strategies for Parallel
  Processing}, 1998.

\bibitem{CoAllocation99}
Karl Czajkowski, Ian Foster, and Carl Kesselman.
\newblock Co-allocation services for computational grids.
\newblock In {\em Proc. 8th IEEE Symp. on High Performance Distributed
  Computing}. IEEE Computer Society Press, 1999.

\bibitem{metampi}
Thomas Eickermann, Helmut Grund, and Jorg Henrichs.
\newblock Performance issues of distributed mpi applications in a german
  gigabit testbed.
\newblock In {\em Proceedings of the 6th European PVM/MPI Users' Group
  Meeting}, September 1999.

\bibitem{mds97}
S.~Fitzgerald, I.~Foster, C.~Kesselman, G.~{von Laszewski}, W.~Smith, and
  S.~Tuecke.
\newblock A directory service for configuring high-performance distributed
  computations.
\newblock In {\em Proc. 6th IEEE Symp. on High Performance Distributed
  Computing}, pages 365--375. IEEE Computer Society Press, 1997.

\bibitem{PhysicsToday}
I.~Foster.
\newblock The grid: A new infrastructure for 21st century science.
\newblock {\em Physics Today}, 54(2), 2002.

\bibitem{mpi-nexus-pc}
I.~Foster, J.~Geisler, W.~Gropp, N.~Karonis, E.~Lusk, G.~Thiruvathukal, and
  S.~Tuecke.
\newblock {A} wide-area implementation of the {M}essage {P}assing {I}nterface.
\newblock {\em Parallel Computing}, 24(12):1735--1749, 1998.

\bibitem{MultiMethodJPDC}
I.~Foster, J.~Geisler, C.~Kesselman, and S.~Tuecke.
\newblock Managing multiple communication methods in high-performance networked
  computing systems.
\newblock {\em Journal of Parallel and Distributed Computing}, 40:35--48, 1997.

\bibitem{GlobusHCW98}
I.~Foster and C.~Kesselman.
\newblock The {G}lobus project: A status report.
\newblock In {\em Proceedings of the Heterogeneous Computing Workshop}, pages
  4--18. IEEE Computer Society Press, 1998.

\bibitem{GridBook}
I.~Foster and C.~Kesselman, editors.
\newblock {\em The Grid: Blueprint for a New Computing Infrastructure}.
\newblock Morgan Kaufmann Publishers, 1999.

\bibitem{GlobusSecurity}
I.~Foster, C.~Kesselman, G.~Tsudik, and S.~Tuecke.
\newblock A security architecture for computational grids.
\newblock Technical report, Mathematics and Computer Science Division, Argonne
  National Laboratory, Argonne, Ill., 1998.

\bibitem{JPDCNexus}
I.~Foster, C.~Kesselman, and S.~Tuecke.
\newblock The {N}exus approach to integrating multithreading and communication.
\newblock {\em Journal of Parallel and Distributed Computing}, 37:70--82, 1996.

\bibitem{Anatomy}
I.~Foster, C.~Kesselman, and S.~Tuecke.
\newblock The anatomy of the grid: Enabling scalable virtual organizations.
\newblock {\em International Journal of High Performance Computing
  Applications}, 15(3):200--222, 2001.

\bibitem{FoxChapterCite}
Geoffrey Fox and Wojtek Furmanski.
\newblock High-performance commodity computing.
\newblock In {\em \cite{GridBook}}, pages 237--255.

\bibitem{pacx}
Edgar Gabriel, Michael Resch, Thomas Beisel, and Rainer Keller.
\newblock Distributed computing in a heterogenous computing environment.
\newblock In {\em Proc. EuroPVMMPI'98}. 1998.

\bibitem{GannonChapterCite}
Dennis Gannon and Andrew Grimshaw.
\newblock Object-based approaches.
\newblock In {\em \cite{GridBook}}, pages 205--236.

\bibitem{pvmbook}
A.~Geist, A.~Beguelin, J.~Dongarra, W.~Jiang, B.~Manchek, and V.~Sunderam.
\newblock {\em {PVM}: {P}arallel Virtual Machine---A User's Guide and Tutorial
  for Network Parallel Computing}.
\newblock MIT Press, 1994.

\bibitem{grim:legion}
A.~S. Grimshaw, W.~A. Wulf, and the Legion~team.
\newblock The {Legion} vision of a worldwide virtual computer.
\newblock {\em Communications of the ACM}, 40(1), January 1997.

\bibitem{mpich}
W.~Gropp, E.~Lusk, N.~Doss, and A.~Skjellum.
\newblock A high-performance, portable implementation of the {MPI} message
  passing interface standard.
\newblock {\em Parallel Computing}, 22:789--828, 1996.

\bibitem{pvmmpi99-mpptest}
William Gropp and Ewing Lusk.
\newblock Reproducible measurements of {MPI} performance characteristics.
\newblock Technical Report ANL/MCS-P755-0699, Mathematics and Computer Science
  Division, Argonne National Laboratory, June 1999.

\bibitem{gropp-lusk-doss-skjellum:mpich}
William Gropp, Ewing Lusk, Nathan Doss, and Anthony Skjellum.
\newblock A high-performance, portable implementation of the {MPI}
  {M}essage-{P}assing {I}nterface standard.
\newblock {\em Parallel Computing}, 22(6):789--828, 1996.

\bibitem{starT}
P.~Husbands and J.C. Hoe.
\newblock {MPI}-{S}tar{T}: Delivering network performance to numerical
  applications.
\newblock In {\em Proceedings of Supercomputing '98}, November 1998.

\bibitem{impi-web}
Interoperable mpi web page.
\newblock \texttt{http://impi.nist.gov}.

\bibitem{optcollops}
N.~Karonis, B.~de~Supinski, I.~Foster, W.~Gropp, E.~Lusk, and J.~Bresnahan.
\newblock Exploiting hierarchy in parallel computer networks to optimize
  collective operation performance.
\newblock In {\em Proceedings of the 14th International Parallel and
  Distributed Processing Symposium}, 2000.

\bibitem{KennedyChapterCite}
Ken Kennedy.
\newblock Compilers, languages, and libraries.
\newblock In {\em \cite{GridBook}}, pages 181--204.

\bibitem{magpie-PC}
T.~Kielmann, H.~E. Bal, S.~Gorlatch, K.~Verstoep, and R.~F.~H. Hofman.
\newblock Network performance-aware collective communication for clustered wide
  area systems.
\newblock {\em Parallel Computing}, 2001.
\newblock accepted for publication.

\bibitem{magpie}
T.~Kielmann, R.F.H. Hofman, H.E. Bal, A.~Plaat, and R.A.F. Bhoedjang.
\newblock {MAGPIE}: {MPI}'s collective communcation operations for clustered
  wide area systems.
\newblock In {\em Proceedings of Supercomputing '98}, November 1998.

\bibitem{stampi}
T.~Kimura and H.~Takemiya.
\newblock Local area metacomputing for multidisciplinary problems: A case study
  for fluid/structure coupled simulation.
\newblock In {\em Proc. Intl. Conf. on Supercomputing}, pages 145--156. 1998.

\bibitem{lam-www}
Collected {LAM} documents.
\newblock World Wide Web.
\newblock {\tt ftp://tbag.osc.edu/pub/lam}.

\bibitem{Olle}
Olle Larsson.
\newblock Implementation and performance analysis of a high-order {CEM}
  algorithm in parallel and distributed environments.
\newblock Master's thesis, University of Houston, 1998.

\bibitem{Condor}
M.~Litzkow, M.~Livny, and M.~Mutka.
\newblock Condor - a hunter of idle workstations.
\newblock In {\em Proc. 8th Intl Conf. on Distributed Computing Systems}, pages
  104--111, 1988.

\bibitem{Quo}
J.~P. Loyall, R.~E. Schantz, J.~A. Zinky, and D.~E. Bakken.
\newblock Specifying and measuring quality of service in distributed object
  systems.
\newblock In {\em Proceedings of the First International Symposium on
  Object-Oriented Real-Time Distributed Computing (ISORC '98)}, 1998.
\newblock Kyoto, Japan.

\bibitem{kumarsc99}
G.~Mahinthakumar, F.~M. Hoffman, W.~W. Hargrove, and N.~Karonis.
\newblock Multivariate geographic clustering in a metacomputing environment
  using globus.
\newblock In {\em Proceedings of Supercomputing '99}. IEEE Computer Society
  Press, 1999.

\bibitem{mpi-forum:journal}
{Message Passing Interface Forum}.
\newblock {MPI}: A message-passing interface standard.
\newblock {\em International Journal of Supercomputer Applications},
  8(3/4):165--414, 1994.

\bibitem{mpi-forum:mpi2-journal}
{Message Passing Interface Forum}.
\newblock {MPI2}: A message passing interface standard.
\newblock {\em International Journal of High Performance Computing
  Applications}, 12(1--2):1--299, 1998.

\bibitem{mpirt-web}
Mpi/rt forum.
\newblock \texttt{http://www.mpirt.org}.

\bibitem{Ninf}
Hidemoto Nakada, Mitsuhisa Sato, and Satoshi Sekiguchi.
\newblock Design and implementations of ninf: towards a global computing
  infrastructure.
\newblock {\em Future Generation Computing Systems}, 15:649--658, 1999.

\bibitem{ncsa-news}
Ncsa press release web page.
\newblock
  \texttt{http://www.ncsa.edu/News/Access/Releases/011211.TeraGrid.html}.

\bibitem{mpich-gq}
A.~Roy, I.~Foster, W.~Gropp, N.~Karonis, V.~Sander, and B.~Toonen.
\newblock {{MPICH-GQ}: Quality-of-Service} for message passing programs.
\newblock In {\em Proceedings of Supercomputing 2000}. IEEE Computer Society
  Press, 2000.

\bibitem{PVMATM2}
T.~Sheehan, W.~Shelton, T.~Pratt, P.~Papadopoulos, P.~LoCascio, and T.~Dunigan.
\newblock Locally self consistent multiple scattering method in a
  geographically distributed linked {MPP} environment.
\newblock {\em Parallel Computing}, 24, 1998.

\bibitem{teragrid}
Teragrid web page.
\newblock \texttt{http://www.teragrid.org}.

\bibitem{webos}
Amin Vahdat, Eshwar Belani, Paul Eastham, Chad Yoshikawa, Thomas Anderson,
  David Culler, and Michael Dahlin.
\newblock {WebOS}: Operating system services for wide area applications.
\newblock In {\em 7th Symposium on High Performance Distributed Computing},
  July 1998.

\end{thebibliography}

Nicholas T. Karonis received a B.S. in finance and a B.S. in computer
science from Northern Illinois University in 1985, an M.S. in computer
science from Northern Illinois University in 1987, and a Ph.D. in 
computer science from Syracuse University in 1992.  He spent 
summers from 1981 to 1991 as a student at Argonne National Laboratory,
where he worked on the p4 message-passing library, automated reasoning,
and genetic sequence alignment.  From 1991 to 1995 he worked on the control
system at Argonne's Advanced Photon Source and from 1995 to 1996 
for the Computing Division at Fermi National Accelerator Laboratory.
Since 1996 he has been an assistant professor of computer science at 
Northern Illinois University and a resident associate guest of Argonne's
Mathematics and Computer Science Division where he has been a member
of the Globus Project.  His current research interests is message- passing 
systems in computational grids.

Brian Toonen received his B.S. in computer science from the University of
Wisconsin Oshkosh in 1993, and his M.S. in computer science from the
University of Wisconsin-Madison in 1997.  He is a senior scientific
programmer with the Mathematics and Computer Science Division at Argonne
National Laboratory.  Brian's research interests include parallel and
distributed computing, operating systems, and networking.  He is currently
working with the MPICH team to create a portable, high-performance
implementation of the MPI-2 standard.  Prior to joining the MPICH team, he
was a senior developer for the Globus Project.

Ian Foster received his B.Sc. (Hons I) at the University of Canterbury
in 1979 and his Ph.D. from Imperial College, London, in 1998. He
is a senior scientist and associate director of the
Mathematics and Computer Science Division at Argonne National
Laboratory, and professor of computer science at the University of
Chicago.  He has published four books and over 150 papers and
technical reports.  He co-leads the Globus Project, which provides
protocols and services used by industrial and academic distributed
computing projects worldwide.  He co-founded the influential Global
Grid Forum and co-edited the book ``The Grid: Blueprint for a New
Computing Infrastructure.''

\end{document}